# CuO chain statistics, charge transfer and $T_c(x)$ dependence in $YBa_2Cu_3O_{6+x}$, $Y_{0.9}(Ca)_{0.1}Ba_2Cu_3O_{6+x}$, and $Y_{0.8}(Ca)_{0.2}Ba_2Cu_3O_{6+x}$

V. M. Matic and N. Dj. Lazarov

*Laboratory of Theoretical Physics,
Institute of Nuclear Sciences "Vinca", 11001 Belgrade, Serbia*

**Abstract**

$x$ dependences of $T_c$ in $YBa_2Cu_3O_{6+x}$ and $Y_{1-b}(Ca)_bBa_2Cu_3O_{6+x}$ ($b=0.1$ and $b=0.2$) have been calculated assuming that the net doping of $CuO_2$ layers is a sum of contributions from CuO chains and from substitution of $Y^{3+}$ by $Ca^{2+}$. We applied the concept of minimal (critical) chain length $l_{cr}$ needed to trigger charge transfer from chains to planes. The model proposed assumes that only a certain part, say $\chi$, of those chain-holes that are created beyond the first $l_{cr}-2$ holes in chains of length $l \geq l_{cr}$, are able to attract electrons from $CuO_2$ bilayer. Our analysis points to the conclusion that parameter $l_{cr}$ should be equal to 4 (four oxygen atoms in a chain), or very close to it (3, or 5). Calculated $x$ dependences of doping, $p(x)$, at constant (room) temperature and for three different substitution levels $b=0$, 0.1, and 0.2, are found to be in excellent agreement with available experimental data. These $p(x)$ dependences are combined with universal (parabolic) phase relation $T_c(p)$ to obtain three $T_c(x)$ dependences that also remarkably correlate with those reported in experiments. The results obtained indicate that in long chains ($x\approx1$) the probability for a chain-hole to capture an electron (expressing hole's ability to become transferred) decreases with the concentration of $3d$ Cu electrons in $CuO_2$ layers, ranking from $\chi\approx40\%(42\%)$ in $YBa_2Cu_3O_{6+x}$, over $\chi\approx36\%$ in $Y_{0.9}(Ca)_{0.1}Ba_2Cu_3O_{6+x}$ to $\chi\approx33\%$ in $Y_{0.8}(Ca)_{0.2}Ba_2Cu_3O_{6+x}$. We estimate that in these three systems the wavelength of charge corrugations in long chains (at $x\approx1$) should be ranking around $\lambda\approx1.38$nm, $\lambda\approx1.25$nm, $\lambda\approx1.20$nm, respectively.

# 1. Introduction

CuO$_2$ planes play a central role in all high-$T_c$ superconducting cuprates (HTSCs) as superconductivity emerges in these materials when a certain fraction of 3$d$ copper electrons, typically ranking between 5% and 27%, is removed from the planes. The missing electrons are commonly referred to as "holes", and they in fact can move throughout the planes acting as charge carriers that make the material conducting and, at low enough temperatures, superconducting. The concentration of holes induced in the CuO$_2$ layers, defined as their number per Cu, is conventionally denoted as "doping" $p$, since in the early stage of high $T_c$ era, in some popular superconductors at the time, the removal of 3$d$ electrons has typically been made by chemical doping through which some interlayer metal ions became replaced by other ions of different valence. Such is the case, for example, in La$_{2-x}$(Sr)$_x$CuO$_4$ in which La$^{3+}$ is substituted by Sr$^{2+}$ in La$_{2-x}$(Sr)$_x$CuO$_4$ which is the process that introduces holes into the layers. The fundamental importance of CuO$_2$ planes is further manifested by the fact that many important physical characteristics, as for example the pseudo gap energy $E_g$ and transition temperature $T_c$, depend on $p$ through universal relations that apply to a wide class of HTSCs [1]. Thus, $p$ dependence of scaled transition temperature $T_c/T_{c,max}$ ($T_{c,max}$ denotes maximal $T_c$) has been found empirically that is a universal, approximately parabolic, function which has its onset, maximum, and termination at $p$=0.05, 0.16 (optimal doping), and 0.27, respectively [2]:

$$T_c(p) = T_{c,\max}\left[1 - 82.6(p - 0.16)^2\right]. \qquad (1)$$

Another doping level to be mentioned is the so-called *critical doping* $p_{crit}$≈0.19, the most frequently defined as the level above which pseudo gap phase disappears [1]. Besides the chemical doping the electrons can also be removed from the planes by introducing oxygen into the material, as in the case of YBa$_2$Cu$_3$O$_{6+x}$. Oxygen is typically incorporated into separate layers in which it orders to form CuO chains that are known, due to the presence of oxygen, to be acting as efficient attractors of electrons. This process has often been identified as "transfer of positive charge (holes)" from chains to plains (that is otherwise equivalent to the transfer of electrons from planes to chains).

The Y$_{1-b}$(Ca)$_b$Ba$_2$Cu$_3$O$_{6+x}$ superconductor ($b$≠0) occupies quite a specific position among high-$T_c$ cuprates because the doping of CuO$_2$ planes is accomplished by combined effect of both mechanisms: substitution of Y$^{3+}$ by Ca$^{2+}$ and addition of oxygen accompanied by formation of CuO chains. As a consequence, the $x$ dependence of $T_c$ reveals utterly different behavior than that of the parent YBa$_2$Cu$_3$O$_{6+x}$ material. Unlike the well known two-plateaus-like shape of $T_c(x)$ of YBa$_2$Cu$_3$O$_{6+x}$, characterized by prominent plateaus at 60K and 90K, the $T_c(x)$ of Y$_{1-b}$(Ca)$_b$Ba$_2$Cu$_3$O$_{6+x}$ for b=0.2 shows only one striking plateau that extends over composition range that nearly matches the regime of extinguished superconductivity in the parent $b$=0 system. Besides of that, in distinction from the YBa$_2$Cu$_3$O$_{6+x}$, whose highly overdoped regime ($p$>0.19) is difficult to attain [2-4], the optimal doping in Y$_{1-b}$(Ca)$_b$Ba$_2$Cu$_3$O$_{6+x}$ ($b$=0.2) is shifted toward lower oxygen concentrations ($x_{opt}$≈0.65) above which $T_c$ shrinks to almost 50% of $T_{c,max}$≈85.5K ($p$≈0.24) [5,6]. Also, $T_c(x)$ for the substitution level $b$=0.1, which we have reproduced here from measured $p(x)$ of Reference [2], shows no any plateau section whatsoever, but reveals a rather parabolic shape with the well pronounced overdoped region. Thus, fully oxygenized materials for $b_1$=0.1 and $b_2$=0.2 have proven themselves particularly suitable for studying various aspects of highly overdoped regime in which normal state electronic correlations associated with the pseudogap phase are nonexistent. The system has therefore been extensively used to study anomalies in the electronic density of states near the Fermi level [7], the emergence of superconductivity in CuO chains induced by proximity effect at $x$ close to 1.0 [8], the doping dependence of phonon renormalization [9], etc.

The aim of this contribution is to explain apparently different behaviors of experimentally observed $T_c(x)$ dependences in three homologous compounds $Y_{1-b}(Ca)_bBa_2Cu_3O_{6+x}$, $b_0$=0, $b_1$=0.1, and $b_2$=0.2 through a common mechanism. We here propose a model in which the total number of holes created in $CuO_2$ planes is a sum of holes that originate from substitution of $Y^{3+}$ by $Ca^{2+}$ and those that come from CuO chains via chains-to-planes charge transfer process, i.e $p=p_b+p_{ch}$. Furthermore, the chain contribution is assumed to originate only from those CuO chains that are longer, or equal to, a certain minimal (critical) chain length needed for charge transfer to take place. The so obtained doping $p(x)$, at constant (room) temperature, when combined with (1) gives $T_c(x)$'s that remarkably correlate with experimental data.

## 2. Charge transfer model

The charge transfer model that we propose here can be concisely introduced by the following expression

$$p = \frac{b_{eff}}{2} + \frac{\chi}{8}\left[n_1\sum_{l=l_{cr}}^{\infty}(l-l_{cr}+1)f_{\alpha 1}(l) + n_2\sum_{l=l_{cr}}^{\infty}(l-l_{cr}+1)f_{\alpha 2}(l)\right], \qquad (2)$$

where the first term stands for $p_b$ (the Ca contribution) and the second corresponds to $p_{ch}$ (the chain contribution). The meaning of quantities introduced in (2) is explained as follows.

**The Ca contribution:** Insofar as each Ca ion introduces one hole and each chain layer supplies holes to two $CuO_2$ planes, one expects to find that the Ca contribution to doping is to be equal to $b/2$, so that, in essence, $b_{eff}$ should stay for $b$. However, experiments of Bertrand *et al.* [4] have shown that at low oxygen content ($x$≈0.02), at which the chain contribution is supposed to be nullified and, consequently, the total doping to be coming only from Ca, values of $p$ at $x$≈0 (that were obtained either from (1) and reported $T_c(x$≈$0)$ (in case of $b$=0.2) [5,6], or measured directly from experiment (the case $b$=0.1) [2]) almost systematically undershoot the expected $b/2$ and correspond to $b_{eff}$ that ranks around 78% of $b$. Therefore, for $b_{eff}$ in (1) we take $b_{eff}$=0.78$b$. It should be mentioned, however, that although the $b_{eff}$ is in fact very well defined in this way, i.e. from the available experimental data [2,5,6], we have nevertheless used it rather as a parameter that should be varied (though, in essence, only very slightly around its expected value $b_{eff}$=0.78$b$) in order to achieve the best possible agreement between the calculated $T_c(x)$'s and the experimental ones.

**The chain contribution:** Although the second term in equation (2), that refers to the chain contribution to doping, has been presented rather extensively in our previous work [24,25], we nevertheless repeat here some of the arguments stated before for the sake of the completeness of the explication. To derive $p_{ch}$ we rely on a widely accepted opinion that copper in chain-plane can be either $Cu^{2+}$, when it is incorporated into the chain interior, or located at chain end (4-fold and 3-fold coordinated Cu, respectively), or $Cu^{1+}$, when it is not included in the chain (2-fold coordinated Cu). This implies that each oxygen atom in a chain, except the first one, has introduced one hole, so that there would be $l$-1 holes created in a chain of length $l$. Given the fact that the state of the chain electronic subsystem, and, consequently, its charge transfer efficiency, should not depend on the history of chain formation, one is free to assume whatever scenario of assembling $l$ oxygen atoms together to form a chain. Thus, instead of to think about a long chain as having been created by merging of two shorter chains, or, for example, by bringing all of $l$ oxygen atoms together at the same instant of time, it is perhaps the most convenient to imagine the chain has been created by adding oxygen one by one, inasmuch as that would allow to follow the evolution of charge transfer process as $l$ increases. A notion that there should exist a minimal (critical) chain length $l_{cr}$ that is required to trigger charge transfer, has naturally emerged through attempts to provide a satisfactory explanation for the existence of 60K plateau in $YBa_2Cu_3O_{6|+x}$ [3,10-12]. According to this reasoning, when $x$ increases beyond $x$=0.5 (stoichiometry of ortho-II phase) oxygen atoms fill empty chain sites (located between CuO chains) in a rather random fashion, therefore creating a large number of isolated oxygen and sporadic short chain fragments. With further increase of oxygen content the $T_c$ remains fairly constant (≈60K) until the chain fragments become sufficiently long to provide additional charge transfer. Such a general scenario is in essence grounded upon two underlying premises: a) the 60K plateau is due to constant doping, $p(x)=const$, over the region of ortho-II phase, and b) there exist a certain threshold chain length $l_{cr}$ so that only chains of length $l≥l_{cr}$ contribute to transfer of holes to $CuO_2$ layers. It should be mentioned that some theoretical studies of internal chain electronic degrees of freedom also provide a backing for the idea of critical chain length [13-15]. The concept of critical (minimal) chain length $l_{cr}$ means that as chain length gradually increases starting from $l$=1 towards $l=l_{cr}$-1, during which process the initial $l_{cr}$-2 chain-holes were created, there is still not enough positive charge for the chain to initiate electron transfer from two $CuO_2$ planes. At $l=l_{cr}$ the charge transfer is triggered so that creation of remaining $l-l_{cr}$+1 chain-holes (as chain length further increases beyond $l_{cr}$) coincides with the ongoing development of charge transfer process. We use to denote the first $l_{cr}$-2 holes as "*the passive holes*", for their creation has not coincided with any transfer of electrons, and the remaining $l-l_{cr}$+1 chain-holes as "*the active holes*", for their creation was accompanied by the arrival of electrons from the planes.

At this point it is worthwhile to make a note that the concept of critical (minimal) chain length is grounded upon a wider underlying idea that a single chain hole is not able to effectively attract an electron, but that only a combined effect of several holes can achieve this goal. This means that one should not expect each active hole to attract an electron for otherwise, had it been so, then the sole chain contribution to doping at $x≈1$ would have been equal to 0.5 (or very close to it). Indeed, at $x$=1 all chains are very long (nearly infinite) and, since there are just a few chain ends in the system, the number of passive holes per Cu is then negligible (regardless of the value of $l_{cr}$) and, to a good approximation, each oxygen can be taken as having introduced one active hole. If all of these holes were transferred the doping would overshoot 0.5 for $b_{eff}/2$. Such a scenario is clearly denied by experiments of Tallon *et al.* [2] and Bertrand *et al.* [6] who obtained that the total doping ($p=p_b+p_{chains}$) is much less than 0.5 ranging around $p(x≈1)≈0.194$, 0.22, 0.24 for $b$=0, 0.1, 0.2, respectively (filled symbols in Figure 1 - the $p(x)$ in the case of $b=0.2$ we have reproduced from reported $T_c(x)$ from the Reference [6] using universal relation (1)). Combining these data with the estimated $b_{eff}$ from $T_c(x)$ at $x\approx 0$ [2,6], one arrives at the conclusion that, in case of long chains (at $x\approx 1$), it is only $\approx 33\%$, 36%, 39% of active chain holes that will succeed in capturing electrons for $b=0.2$, 0.1, 0, respectively. At this point we introduce a new quantity, $\chi$, defined as the number of effectively attracted electrons per active hole. Furthermore, we also introduce a new assumption that the active hole efficiency $\chi$ to capture $CuO_2$ electrons, as determined by $\chi\approx 0.33$, 0.36, 0.40 for $b=0.2$, 0.1, 0, respectively, maintains the same value not only in case of long chains that prevail at $x\approx 1$, but also for fragmented chains that dominate in underdoped regime. In other words, we postulate $\chi$ is of the same value in all chains, ranging from $l=l_{cr}$ to $l=\infty$, so that the number of transferred holes from a chain of length $l\geq l_{cr}$ turns to be equal to $\chi(l-l_{cr}+1)$. It is interesting to note that although in the case of $YBa_2Cu_3O_{6|+x}$ we estimated $\chi$ from experimental data at $x\approx 1$, the estimated value $\chi\approx 39$-40% agrees very well with what one would expect at ortho-II stoichiometry $x\approx 0.5$. Since at $x\approx 0.5$ long chains alternate along $a$-axes with rows of empty sites, the concentration of active holes $h_a$, defined as their total number per Cu, cannot exceed 0.5, given the fact that the concentration of passive holes is here negligible as at $x\approx 1$. The $h_a$ can really be only slightly less than 0.5 due to thermally activated chain fragmentation (as it will become apparent from sections 3 and 4, $h_a$ maintains its value (for example $h_a\approx const=0.48$) as at $x\approx 0.5$, so in the whole interval of the 60K plateau). On the other hand, from (1) it follows that 60K of $T_c$ corresponds to doping level $p\approx 0.0945$, which, from $p=(\chi/2)h_a$, makes it obvious that $\chi$ again falls at some point around 39% (40%). We are therefore not of opinion that 60K plateau is due to doping level $p\approx 0.125=const$ as it has been inferred in some studies [3]. Since the ortho-II phase is a highly non homogeneous it is difficult to convincingly measure the distances between the atoms and to evaluate accurately the bond valence sums, so we are of opinion that the so-called "1/8 dip" in $T_c(p)$ dependence, that has been reported exclusively in $YBa_2Cu_3O_{6|+x}$ superconductor, is probably to be assigned to imperfect measurement of doping in the region of the 60K plateau.

Out of the way the parameter $\chi$ has been introduced above, it follows that it reflects the capability of holes to attract electrons. Generally, one would expect that the hole capability, as expressed by the value of $\chi$, should be affected by physical conditions in the nearest chain coordination. Aside from a certain coupling between chains and planes, chances for a hole to capture an electron should larger as more $3d$ copper electrons are available in the planes immediately above (below) the chain. In other words, $\chi$ should be increasing with concentration, $\rho_e$, of $3d$ electrons, but at this stage we can only speculate whether it would be proportional to $\rho_e$ (for single layered cuprates), or to $\rho_e^2$ (double layered). Given the fact that introduction of Ca takes away more electrons from $CuO_2$ planes than in the parent $YBa_2Cu_3O_{6+x}$ compound, and therefore additionally reduces $\rho_e$, it is not surprising that $\chi$ decreases with substitution level $b$.

Let $N_{Cu}$, $n$, and $f(l)$ denote the number of Cu atoms in a chain plane, the fraction of the 3-fold coordinated Cu (residing on chain ends) and the fraction of chains of length $l$, respectively. The number of holes transferred to one $CuO_2$ plane is then equal to $P=\sum_{l=l_{cr}}^{\infty}(\chi/2)(l-l_{cr}+1)f(l)(n/2)N_{Cu}$. It should be noted that although, unfortunately, no systematic experimental study on structural phase diagram of $Y_{1-b}(Ca)_bBa_2Cu_3O_{6+x}$ ($b\neq 0$) system has been made by now, the ortho-II phase has nevertheless been reported at

$x \approx 0.5$ [16]. On the other hand, in the light of the fact that superstructure reflections of other orthorombic phases for example, orthi-III, ortho-IV, etc., were found to be of considerably reduced intensity [17], it has often been inferred that these phases emerge only in small patches embedded in large domains of the main phases [18]. Besides of that it is very well known that in $YBa_2Cu_3O_{6+x}$ system no other chain phases except ortho-II was reported at $x>0.62$, indicating that it is this phase that lies at the root of the 60K plateau. So, we used here the two dimensional asymmetric next-to-nearest neighbor Ising (ASYNNNI) model to describe oxygen ordering processes in chain plane, for the model has long been known that stabilizes both major phases, ortho-I-and ortho-II, as its ground states [19]. In the case of ortho-II phase, when oxygen chain sites split into two interpenetrating square sublattices commonly known as $\alpha_1$ and $\alpha_2$, the chain contribution to doping of a single $CuO_2$ layer $p=P/N_{Cu}$, attains the form of the second term in equation (2). The quantities $n_1$, $n_2$ and $f_{\alpha 1}(l)$, $f_{\alpha 2}(l)$ stand for the fractions of 3-fold coordinated Cu and length distribution functions of CuO chains on sublattices $\alpha_1$ and $\alpha_2$, respectively.

### 3. A brief remarks on the basics of the ASYNNNI model thermodynamics

Ground state of the ASYNNNI model in the composition interval $0.5 \leq x \leq 1$ is very well known: it consists of either completely occupied, or completely empty, rows of oxygen sites that lie parallel to the *b*-axes. As the rows are connected by Coulomb $V_3>0$ bonds it seems as if the ground state configurations have been generated by $V_3$ coupled one dimensional (1*d*) Ising model. Such visual resemblance might lead someone to conclude that, at nonzero (low) temperatures, the ASYNNNI model can be to a good approximation mapped onto the $V_3$-coupled 1*d* Ising chain, i.e. that the basic thermodynamic functions of the two models are homologous at $T \approx 0$. However, the conclusion of this kind is apparently incorrect, as it can be deduced by analyzing microscopic features of exited states of the ASYNNNI model. The first exited state is generated when one of completely occupied rows is split into two parts and all oxygen atoms from one part subsequently relocated to an unoccupied row. Such a transformation (as shown, for example, in Figure 2 of reference [20]) produces energy change $\Delta E=4|V_2|$ and is accompanied by creation of two unlike $V_2<0$ bonds (copper mediated superexchange next-to-nearest neighbor O-O interaction). Other exited states are obtained in a similar fashion, so that the energy of the *k-th* exited state is determined by

$$E_k = E_o(x) + 2N_{V_2}^{+,-}|V_2| \quad , \tag{3}$$

where $N_{V_2}^{+,-}=2k$ denotes the number of unlike $V_2$ bonds [21]. From the above equation it follows

$$\frac{\langle E \rangle}{N} = \frac{E_0(x)}{N} + |V_2|n(x,T) \quad , \tag{4}$$

where *n* denotes the fraction of 3-fold coordinated Cu ions, $N=2N_{Cu}$ is the number of oxygen sites in the basal (chain) plane, $N_{Cu}$ is the number of Cu ions, and $<E>$ is statistically averaged the total energy of the ASYNNNI model.

Let $\{\sigma_i\}$ denotes a particular configuration of Ising spins and let $g(\{\sigma_i\})$ be a certain statistical quantity (for example, *g* can be a spin-spin correlation function between Ising spins at arbitrary positions, the deviation of the total energy from the energy of the ground state, $\Delta E=E-E_0(x)$, or corresponding expression for entropy). Given the fact that, according to (1), the Boltzman weighting factor splits into product of two terms, it follows that, at $T\approx 0$ the equilibrium value of $g$ can be expressed in the following way

$$g = \frac{\sum_{\{\sigma_i\}} g(\{\sigma_i\}) e^{-\frac{E_0(x)}{k_B T}} e^{-\frac{2|V_2|N_{V2}^{+,-}(\{\sigma_i\})}{k_B T}}}{\sum_{\{\sigma_i\}} e^{-\frac{E_0(x)}{k_B T}} e^{-\frac{2|V_2|N_{V2}^{+,-}(\{\sigma_i\})}{k_B T}}}, \tag{5}$$

Since, in the above expression, the $V_3$ interaction participates only in $E_o(x)$ it follows that no one of these quantities ($\Delta E$, entropy per site, pair correlation functions, etc.) does not depend on $V_3$ in the regime of low temperatures. Therefore, it can be stated that the whole of the thermodynamics of the ASYNNNI model does not practically depend on the magnitude of $V_3$ at $T\approx 0$ (aside from the fact that only the total energy $E$ depends on $V_3$ through the ground state term), but can depend only on $V_2<0$ (the nearest neighbor interaction $V_1$ acts only as a scaling factor for the temperature).

An extended analysis have shown that at low $T$ the fractions $n_1(x,T)$ and $n_2(x,T)$ of 3-fold copper (where $n=(n_1+n_2)/2$) depend on $V_2$, and on $x$ and $T$, through the products that separate variables $x$ and $T$ according to

$$n_1 \cong \varphi_1(x)\exp(-2|V_2|/k_B T), \quad n_2 \cong \varphi_2(x)\exp(-2|V_2|/k_B T). \tag{6}$$

and that sublattice occupations $x_1$ and $x_2$ depend only on $x$, but not on $T$ [21]. Thus, for example, in the three five-point plus three four-point basic cluster approximation of the cluster variation method (CVM), these four x dependences have the following analytic form

$$\varphi_m(x) = 2(2x-2)\sqrt{\frac{1-x\mp(\eta/2)}{x\pm(\eta/2)}}, \quad x_m(x) = x\pm\eta/2, \quad m=1,2, \tag{7}$$

where upper sign corresponds to $m=1$ and lower $m=2$ [21]. The order parameter between ortho-II and ortho-I phases, $\eta = x_1 - x_2$, is given by

$$\eta(x) = 2x\sqrt{\frac{3-4x}{4x-1}}. \tag{8}$$

The very fact that at $T\approx 0$ the energy $\Delta E$ depends on $T$ through the factor $\exp(-2|V_2|/k_B T)$ is manifestation of a more general behavior

$$\Delta E \propto \left[1 - \tanh\left(\frac{|V_2|}{k_B T}\right)\right], \tag{9}$$

which is attributable to the $V_2$ coupled linear Ising chain in zero external field. We are therefore to bring about the conclusion that at low temperatures the ASYNNNI model behaves as $V_2$ coupled 1d Ising model rather than the $V_3$ coupled 1d Ising model. This means that at $T\approx 0$ no one row of oxygen chain sites is either completely occupied or completely empty, but along each row the CuO chain fragments alternate with empty fragments. As $T$ approaches absolute zero, for a given $x$, both full and empty chain fragments become virtually all longer, but oxygen occupancy remains constant, i.e. independent on $T$, being equal either to $x_1(x)$, or to $x_2(x)$, depending onto what sublattice ($\alpha_1$, or $\alpha_2$) the particular row belongs to. Furthermore, as $T\rightarrow 0$, the fractions $n_1$ and $n_2$ of 3-fold coordinated Cu, which determine number of chain ends, vanish as $\propto \exp(-2|V_2|/k_B T)$, as well as the energy deviation from the ground state energy, $\Delta E=E-E_0(x)$.

For given $x$ (0.5<$x$<1), fluctuations of energy virtually tend to infinity, accompanied by the temperature dependence of specific heat $c \propto (1/T^2)\exp(-2|V_2|/k_BT)$, exactly as in the case of $V_2$ coupled linear chain Ising model in zero external field. In addition, the so-called non-ordering susceptibility $\partial x/\partial \mu$ ($\mu$ stands for the chemical potential) reveals an essential singularity of the form $\partial x/\partial \mu \approx (1/T)\exp(2|V_2|/k_BT)$ [22], that is clearly a signature of $V_2$-coupled 1$d$ Ising chain nature of the ASYNNNI model at low temperatures. We are therefore about to conclude that the ASYNNNI model undergoes the second order phase transition at $T=0$ axis of $(x,T)$ space of the kind that is known to occur in zero field linear Ising chain, with the role of the *NN* interaction of the later being played by the *NNN* $V_2$ interaction of the former.

Although the magnitude of $V_3$ interaction becomes irrelevant at temperatures that are sufficiently low, its existence in the ASYNNNI model, as manifested through its positive (repulsive) nature, has nevertheless important implications. At first, it is the $V_3$ that stabilizes ortho-II phase by the virtue of its repulsiveness. Secondly, since increase of $V_3$ lowers the ground state energy at $x=0.5$, it means that the magnitude of $V_3$ may well affect the location of the top of ortho-II phase along the $T$-axis of $(x,T)$ space (we denote it by $T_{OII}$). Besides of that, the magnitude of $V_3$ also affects the upper limit, $T_u$, of low temperature region $0<T<T_u$ within which the ASYNNNI model turns to its $V_3$-independence, i.e. to its $V_2$-coupled 1$d$ Ising chain model nature. The temperature $T_u$ can be determined, for example, as the highest temperature at which expressions (6)-(8) still fairly accurately reproduce equilibrium values of $n_1$, $n_2$, $x_1$, and $x_2$ (obtained, for example, either by Monte Carlo simulations, or by CVM calculations). Our analysis shows that the $T_u$ generally falls at some point not too far below the top of ortho-II phase and that the interval $0<T<T_u$ generally encompasses the room temperature. In this way, CuO chain formation in actual samples of $Y_{1-b}(Ca)_bBa_2Cu_3O_{6+x}$ ($b_1=0.1$; $b=0.2$) and $YBa_2Cu_3O_{6+x}$ compounds can be taken as being fairly well described by the low-temperature statistics of the ASYNNNI model. Besides of that, as our analysis has indicated (not shown here), the conclusion of this kind applies as well to other models generated by corresponding extensions of the ASYNNNI model (designed in the way to enable stabilization of other chain structures - commonly known as OIII, OIV, etc).

**Length distribution of CuO chains:** Consider a row of $N_b$ oxygen chain sites that is parallel to $b$-axis ($N_b$ is a large number) containing a fixed number of oxygen atoms, say $xN_b$ (in case of ortho-I phase $x$ is then equal to overall oxygen concentration in basal plane). These oxygen atoms are generally divided into a certain number, $N_{V2}^{+;-}/2=(n/2)N_b$, of CuO chains. In principle, the number of chains may vary from 1 to $N_b/2$ (in the later case all oxygen atome would be isolated). Therefore, for given $x$, configuration space $\Omega(x)$ of the whole lattice can be expanded in a sum

$$\Omega(x) = \sum_n \Omega(n) \ , \tag{10}$$

where $\Omega(n)$ stands for the number of configurations with a fixed number of chains. In the above equation it is implicitly assumed that the number of chain ends (i.e. the number of unlike $V_2$ bonds) is homogeneously distributed over all rows of oxygen sites. Given the fact that, in the low temperature regime $0<T<T_u$, the quantities $n$ and $E$ are connected by one-to-one relation (4) (or (3)), it follows that the above sum can be rewritten as

$$\Omega(x) = \sum_E \Omega(E) \ . \tag{11}$$

In statistical mechanics it is very well known theorem that, in the thermodynamic limit $N \rightarrow \infty$, the logarithm of $\Omega(x)$ can be replaced by the logarithm of the leading term, $\overline{E}$, as long as fluctuations of energy around $E = \overline{E}$ are not too large. Therefore, away of critical region $\Omega(x)$ can be to a good approximation replaced with $\Omega(\overline{E})$, i.e. with $\Omega(\overline{n})$, where $\overline{n}$ is linked to $\overline{E}$ through equation (4). This implies that, e.g. in ortho-I phase, a samples with fixed $x$ will practically have fixed number of chains (i.e. fixed number of chain ends) as long as the system is kept apart enough from the critical regime.

Since each CuO chain has two ends, we can think of them as that one would be "positive" (oriented towards the positive side of $b$-axes) and, consequently, the other one would be "negative" chain end. Among $xN_b$ oxygen atoms on a particular row there will be $(n/2)N_b$ of them that reside on positive chain ends. Therefore, the probability for an oxygen atom to be located at the positive chain end is equal to $\omega = n/2x = (l_{av})^{-1}$ ($l_{av}$ denotes the average chain length), and, consequently, the probability for an oxygen atom to be lying either at the negative chain end, or within the interior of the chain, is accordingly equal to $1-\omega$. Assuming that a chain has been constructed by adding oxygen one by one, starting from its negative end, one arrives at the conclusion that probability for the chain to attain length $l$ is equal to $\omega(1-\omega)^{l-1}$. An extended analysis have shown that basically the same conclusion holds for each of two oxygen-site sublattices $\alpha_1$ and $\alpha_2$ so that length distributions of CuO chains are given by [20]

$$f_{\alpha i}(l) = \frac{1}{l_{av,\alpha i}} \left(1 - \frac{1}{l_{av,\alpha i}}\right)^{l-1}, \quad i = 1, 2 \quad , \tag{12}$$

and also that the above expressions hold fairly well not only at low temperatures, but as well accurately at temperatures that are very high indeed, extending up to $\approx 1800K$ (in the regime of ortho-I phase [23]). The only exception where (12) is not reproducing calculated (equilibrium) values of $f_{\alpha i}(l)$'s that accurately is connected, as expected from (11), with the critical regime of ortho-I-to-ortho-II second order phase transition at $x>0.5$. However, such a shortcoming of (12), as manifested by a certain departure of $ln[f_{\alpha i}(l)]$ dependences from the straight-line behavior, has only been observed in a relatively narrow region around critical ortho-I-to-ortho-II phase transition line, of the width generally not exceeding $\Delta x \approx 0.07$ [23]. Furthermore, when $f_{\alpha i}(l)$'s, that are calculated directly by Monte Carlo method, are inserted into (2) no detectable effect is observed in obtained $p(x)$'s at $T=const$ comparing to the case when (12) is combined with (2), implying that the departures of length distributions from (12), around the critical points of ortho-I-to-ortho-II transition, are somehow compensated by summations in (2) [25]. In view of this, it follows that the equation (2) can also be used as transformed into a closed analytic expression, as it has been done in reference [24].

Taking into account equations (6-8), it is obvious, when $f_{\alpha i}(l)$'s are inserted into (2), that the concentration of active chain holes $h_a$ (as defined by $p_{ch}=(\chi/2)h_a$) is a function that depends upon three variables: $x$, $T$, and $l_{cr}$. Besides of that, the $h_a$ depends on the input interaction parameters $V_1$, $V_2$, and $V_3$ of the ASYNNNI model in the way that it in fact depends only upon $V_2$, which in turn emerges only as being coupled to $T$ through the ratio $V_2/k_BT$. Furthermore, as it has been shown in reference [25], at $T=const$ the different functions $h_a(l_{cr})$, that correspond to values of oxygen concentration, $x_i$, spanning over the region of ortho-II phase, all intersect at a single, well defined, value of $l_{cr}$. We use to denote this value as the *optimal critical length* for the particular temperature, $l_{opt,cr}(T)$, for it is the value of $l_{cr}$ for which the $h_a(x)$ and, consequently, the $p_{ch}(x)$ maintain at a constant value over the region of ortho-II phase at the considered $T=const$. Obviously, the constant value of $h_a$, i.e. $h_a(x)=h_a(x,T,l_{opt,cr}(T))=const$ at $T=const$, cannot exceed 0.5; it typically gets closer to 0.5 as the temperature is lowered, but it generally does not fall to much below 0.5 even at pretty high temperatures (extending up to those that lie not too far below the top of ortho-II phase). As it will be shown in the following section, the case $l_{cr}=l_{opt,cr}(T)$ refers to the two plateaus $T_c(x)$ phenomenon in Ca free YBCO system, i.e. when $l_{cr}=l_{opt,cr}(T)$ is applied to (2) it accounts for both 60K amd 90K plateaus of $YBa_2Cu_3O_{6|+x}$. On the other hand, if, at a given $T=const$, the value of parameter $l_{cr}$ is taken to be less than the corresponding $l_{opt,cr}(T)$, then $h_a(x)=h_a(x,T,l_{cr})$ at $T=const$ is a monotonically increasing function in the region of ortho-II phase [25] (in fact, in this case, $h_a(x)$ monotonically increases over the whole range $0<x<1$ of oxygen composition; the same also applies for $p_{ch}(x)=(\chi/2)h_a(x)$). From what follows it will become apparent that the case $l_{cr}<l_{opt,cr}(T)$ applies to Ca doped YBCO, where the different levels of Ca substitution are taken into account only through $b_{eff}=0.78b$ in the equation (2). The third case, when $l_{cr}>l_{opt,cr}(T)$, is not of particular interest here; we just make a mote that in this case $h_a(x)$ reveals a maximum at $x\approx 0.5$ (stoichiometry of ortho-II phase) that is followed by a minimum at $x=0.70\div 0.80$, before it turns to be increasing to eventually become equal to 1, at $x=1$.

In conclusion to this section, we briefly recapitulate the most relevant properties of the ASYNNNI model:

A. The isomorphism that exists between the ASYNNNI model at low temperatures, $0<T<T_u$, and the $V_2$-coupled $1d$ Ising chain model in zero external field, leads to $x$-dependences of chain fragmentation, as determined by $x_1(x)$, $x_2(x)$, and $\varphi_1(x)$, $\varphi_2(x)$ (equations (6-8), to be independent as on the constant temperature, $T<T_u$, at which calculations are made, so as on the input interaction parameters $V_1>0$, $V_2<0$, $V_3>0$, that define the ASYNNNI model. In this view, the expressions (6-8) can be taken as *universal characteristics* of the ASYNNNI model.

B. As a consequence of what has been stated in A, the $h_a(x)$ at $T=const$ reveals either constant section on the right hand side of $x\approx 0.5$ followed by an increase upon entering the regime of ortho-I phase (which takes place if $l_{cr}=l_{opt,cr}(T)$), or monotonically increases over the whole range of oxygen composition $0<x<1$ (when $l_{cr}<l_{opt,cr}(T)$).

## 4. Results of the calculations

In Figure 1 we show experimental values of the $p(x)$ dependences in $Y_{1-b}(Ca)_bBa_2Cu_3O_{6+x}$ cuprate family that we scanned from references [2] ($b=0$ and $b=0.1$, shown by black circles and triangles, respectively) and [5,6] ($b=0.2$, shown by black squares). While, in case of $b=0$ and $b=0.1$, the values of $p(x)$'s were obtained in a direct way, evaluating bond valence sums as determined from measured interatomic distances [2], in case of $b=0.2$ the corresponding $p(x)$ values were derived by ourselves applying universal relation (1), in the reverse direction, to measured $T_c(x)$ [5,6]. In figure 2 are shown the corresponding $T_c(x)$ dependences (with the respective black symbols) obtained either by combining (1) and reported $p(x)$'s (for $b=0$ and $b=0.1$) [2], or scanned directly from [5,6] ($b=0.2$). From Figure 1 it can be seen that for $b=0$ the experimental data on $p(x)$ are missing around $x \approx 0.5$, which is due to the fact that the doping is not convincingly measurable in ortho-II phase [2,10]. However, the two points that are nevertheless available at $x \approx 0.6$ have already faded slightly below $p=0.10$, thus clearly suggesting that it is highly unlikely that the 60K of $T_c$, that approximately corresponds to $x$ lying between $x \approx 0.5$ and $x \approx 0.6$, would be connected with the doping level $p=0.125$ and to the related "1/8 dip" phenomenon of $T_c$ versus $p$ relation (incidentally, this mild depression of $T_c$ from universal $T_c(p)$ behavior (1) has been detected only in $YBa_2Cu_3O_{6|+x}$ system). Accordingly, we believe that the "dip at 1/8" is an artifact of unconvincingly measurable doping in the of ortho-II phase, which is otherwise highly pronounced in Ca-free $YBa_2Cu_3O_{6|+x}$ system. We therefore give an advantage to the universality of (1) against weakly manifested depression of $T_c(p)$ at $p=1/8$, and express our strong opinion that the 60K plateau in $YBa_2Cu_3O_{6|+x}$ is due to the constant doping level at $p=0.0946$ ($=const$) over the regime of ortho-II phase, as it is illustrated in calculated $p(x)$ dependence that is shown in Figure 1 by open circles (the details of the calculation will be given below).

To calculate $p(x)$ from (2) one needs to know, aside from the parameters $b_{eff}$ and $\chi$, the critical chain length $l_{cr}$ and the scaled temperature $\tau = k_B T/V_1$ that the room temperature is to be referred to. Furthermore, the interactions $V_1$, $V_2$, and $V_3$ of the ASYNNNI model should also be known. Unfortunately, it seems that a general agreement about what the magnitudes of these interactions would be equal to, even in the case of extensively studied $YBa_2Cu_3O_{6|+x}$, has not been achieved yet, although there is some consensus that the nearest neighbor O-O interaction $V_1$ should be ranking around 6.7-6.9mRy [18,26]. Even if this estimate of $V_1$ is assumed confident, it would be difficult to convincingly evaluate $V_2$ and $V_3$ from, for example, the temperature $T_{OII}$ that corresponds to the top of ortho-II phase, not only because a single parameter ($T_{OII}$) is not enough to determine two parameters, but also because nobody has decisively resolved so far what would be the exact form of the $T_{OII}(V_1,V_2)$ function. Given the above facts, we here used interactions as suggested in reference [26] for all three systems $YBa_2Cu_3O_{6|+x}$, $Y_{0.9}(Ca)_{0.1}Ba_2Cu_3O_{6+x}$, and $Y_{0.8}(Ca)_{0.2}Ba_2Cu_3O_{6+x}$, since our primary goal was to extract as more conclusions as possible, aimed to be, at the final instant, independent on the actual magnitudes of interactions.

To estimate $\tau_{RT}$ (i.e. the $\tau$ that refers to room temperature) for the $YBa_2Cu_3O_{6|+x}$ system, we applied the reasoning as in references [24] and [25]: Inasmuch as the $V_1 \approx 6.7$mRy sets the scaling between $T$ and $\tau$ so that $\Delta\tau \approx 0.1$ corresponds to $\Delta T \approx 100$K, and since the experimentally obtained $T_{OII}$ ranges around $\approx 125°C$, together with the theoretically obtained $\tau_{OII}$ that is lying around $\approx 0.58$ (for the interactions parameters from ref. [26]), we thus arrived at $\tau_{RT}=0.45$ as a fairly reliable estimate of the room temperature in the Ca free system. As shown in [25], that puts $l_{opt,cr}(\tau=0.45)$ at some point between 4 and 5 (oxygen atoms in the chain). We applied Monte Carlo method using single-spin-flip Glauber dynamics to calculate doping (2), at $\tau=const$ and for $l_{cr}=4$, for a set of oxygen compositions $x$ that span the whole interval $0<x<1$ (in the Glauner dynamics scheme the oxygen concentration $x$ (i.e. $x_1$ and $x_2$) is a function of temperature and chemical potential). The system that the calculations were made on typically consisted of 200X200 of oxygen α (chain) sites, although in some cases the systems with 200X600, and even with 200X800, sites were used. At each calculated point of the $(x,T)$ space (i.e.

of the ($x,\tau$) space) we have been using to run two parallel approaches in obtaining doping: a) The first one relied upon calculating the first three hundred terms, by obtaining the chain length distributions $f_{\alpha1}(l)$ and $f_{\alpha2}(l)$ for $l=1$ to $l=300$, to evaluate the sums in (2), and b) the second one was based on using the integrated form of (2), that was derived by inserting (12) into (2), in which case only the four quantities $x_1$, $x_2$, $n_1$, and $n_2$ were required. The length distributions were obtained by obtaining ratios $N_{\alpha1}(l)/N_{\alpha1}$ and $N_{\alpha2}(l)/N_{\alpha1}$ in each MC step ($N_{\alpha1}(l)$ and $N_{\alpha2}(l)$ stand for the number of chains of the same length $l$, while $N_{\alpha1}$ and $N_{\alpha2}$ are the total numbers of chains, on corresponding sublattice of oxygen chain sites), and these were subsequently equilibrated through the MC process and eventually identified with $f_{\alpha1}(l)$ and $f_{\alpha2}(l)$ [25]. At all calculated points ($x,\tau$), including even those at the very ortho-I-to-ortho-II second order phase transition (at $x>0.5$), the values of doping obtained through these two different approaches were practically indistinguishable, signifying that indeed the departure of $\Omega(x)$ from $\Omega(\overline{E})$ (equation (11)) that occurs in the critical regime, as manifested by departure of $f_{\alpha i}(l)$'s ($i=1,2$) from (12), has been compensated in a certain way by summations in (2). The so calculated $p(x)$, for $\chi=0.414$ ($b_{eff}=0$) is shown by open circles in Figure 1. It can be seen that calculated $p(x)$ correlates fairly well with experimentally obtained values of $p(x)$ (filled circles) and, furthermore, it reveals a constant section at the doping level slightly less than ≈0.1 (it is in fact $p≈0.0946=const$) below $x≈0.6$ and terminating at $x≈0.5$, which is what we believe that would have been obtained in experiments had it been that the doping were measurable in a convincing way in ortho-II phase. These results were used to obtain the corresponding $T_c(x)$ that is shown in Figure 2, using the same respective symbols. The apparent coordination between calculated $T_c(x)$, that displays clearly distinguished plateaus at 60K and 90K, with the experimental result (shown by a solid line that is scanned from reference [4]), points to the conclusion that indeed the first plateau is due to the constant foping section at $p≈0.095$ over the oxygen composition range of the ortho-II phase, while the 90K quasi-plateau is connected with the universal $T_c$ versus $p$ relation (1). It should be also mentioned that if a lower value than $\tau=0.45$ were assigned to $\tau_{RT}$, its $l_{opt,cr}(\tau)$ would be greater, but the corresponding $p(x)$ would have a more pronounced plateau, stretching from $x≈0.5$ to greater values of $x$ at it's the right edge, so that it would approach the $x=1.0$ axes by a sharper angle (this means that $p(x)$ of a greater value of the parameter $l_{cr}=l_{opt,cr}(\tau)$ would lie below that of the smaller $l_{cr}$, at approximatelly $x>0.7$, which is understandable in view of the fact that as greater the $l_{cr}$ as more terms are to be subtracted from sums in (2)). This implies that such $p(x)$ would cross the optimal doping level at the value of $x$ that is more closer to $x=1$ than $x≈0.92$ and, consequently, the 90K quasi-plateau would turn out to be less pronounced (in fact, it would be shorter and it would look like as if it is anomalously pinned to the $x=1.0$ axes – much the same as it is shown in Figure 3). That is why we are pretty sure that in YBa$_2$Cu$_3$O$_{6|+x}$ system the parameter $l_{cr}$ is hardly to be greater than 5.

As to the Y$_{0.9}$(Ca)$_{0.1}$Ba$_2$Cu$_3$O$_{6+x}$ and Y$_{0.8}$(Ca)$_{0.2}$Ba$_2$Cu$_3$O$_{6+x}$ systems, from Figure (1) it can be noted that their $p$ versus $x$ dependences show no any horizontal sections whatsoever, but instead display a rather monotonic increase over the whole range of oxygen composition 0<$x$<1. Out of what has been discussed in section 3 it follows that such behavior points to the quantities $\tau_{RT}$ and $l_{cr}$ as they should be ascribed such values to ensure the relation $l_{cr}<l_{opt,cr}(\tau_{RT})$. There is no a particular reason to a priori believe the $l_{cr}$ parameter should be of universal value in all three systems Y$_{1-b}$(Ca)$_b$Ba$_2$Cu$_3$O$_{6+x}$, $b=0$,

0.1, and 0.2, since the ability of a chain to trigger the charge transfer might well be influenced by altered charge distributions in its adjacency. However, in $Y_{0.9}(Ca)_{0.1}Ba_2Cu_3O_{6+x}$ and $Y_{0.8}(Ca)_{0.2}Ba_2Cu_3O_{6+x}$ it nevertheless seems unlikely that the value of $l_{cr}$ would be significantly different from that one of the $YBa_2Cu_3O_{6|+x}$ compound, given the fact that there is essentially the same basic mechanism, in otherwise homologous systems, that is lying behind the chain's ability to initiate the transfer of electrons (although the exact details of this mechanism are not known yet). Therefore, the most reasonable values of $l_{cr}$ in both $Y_{0.9}(Ca)_{0.1}Ba_2Cu_3O_{6+x}$ and $Y_{0.8}(Ca)_{0.2}Ba_2Cu_3O_{6+x}$ are likely ranking around 3, 4, or 5. Although we are not convinced in advance about universal character of the parameter $l_{cr}$ (i.e. about its *b*-independence) we are neither oposing to such an attitude, since there has not been yet a compelling evidence for either possibility. We therefore proceed by taking $l_{cr}=4$ as in bare $YBa_2Cu_3O_{6|+x}$, motivated primarily by an intention to investigate whether the same value of $l_{cr}$ accounts for all three YBCO systems (i.e. the systems with *b*=0, *b*=0.1 and *b*=0.2, respectively). As to the $\tau_{RT}$ and in-plane O-O interactions $V_1$, $V_2$, and $V_3$ it should be recalled that in fact only the copper mediated $V_2$ interaction is of importance, and that it thereto emerges only as coupled to $\tau$ through $\xi=V_2/k_BT=(V_2/V_1)/\tau$ as shown by (5-9). Thus, the possibly altered interactions $V_1$, $V_2$, and $V_3$, and thence the $\tau_{RT}$, in Ca-doped YBCO will affect the wholeness of the ASYNNNI model statistics only through this ratio. Inasmuch as in $Y_{0.9}(Ca)_{0.1}Ba_2Cu_3O_{6+x}$ and $Y_{0.8}(Ca)_{0.2}Ba_2Cu_3O_{6+x}$ systems there have been no reports thus far about what the magnitudes of the three O-O interactions might be equal to, neither on the structural phase diagram that would include a precise dislocation of the main, ortho-II and ortho-I, phases (and, consequently, the temperature $\tau_{OII}$ of the top of the ortho-II phase), it is difficult to make a reliable estimation for the $\tau_{RT}$ (it should be noted, however, that clear signals of both major phases in *b*≠0 systems have nevertheless been registered in experiments [16]). Considering such a development, we have made a choose to use interactions of the same magnitudes as given in reference [26], mainly with a view to examine to what degree the combined effect of both the doping expression (2) and the statistics of the ASYNNNI model of O-ordering in planes account for $T_c(x)$ characteristics in Ca-doped YBCO systems (with *b*=0.1 and *b*=0.2). We therefore attached the room temperature to a lower value of scaled temperature, $\tau_{RT}=0.35$, than in the Ca-free system (in which it was $\tau_{RT}=0.45$), basically because at $\tau=const=0.35$ the $l_{opt,cr}(\tau)$ falls at some point between 8 and 9 [25], which stands well beyond $l_{cr}=4$ that is adopted here to be used in (2). Thus, the condition $l_{cr}< l_{opt,cr}(\tau)$ for monotonically increasing $p(x)$ is now fulfilled at $\tau=0.35=const$. Taking into account that at the relevant temperature interval $0<T<T_u$ only the magnitude of $V_2$ really matters (but not of the $V_3$ and $V_1$) and that $\tau$ and $V_2$ enter into calculations as coupled through $\approx V_2/\tau$, it might be stated that our estimate $\tau_{RT}=0.35$ fixes not only scaling between the room temperature and the scaled temperature $\tau$, but also includes the effect of possibly altered $V_1$, $V_2$, and $V_3$ in Ca-doped YBCO systems.

The chain length distributions $f_{\alpha 1}(l)$ and $f_{\alpha 2}(l)$ were determined in the following way: In each MC step we counted the total numbers of chains $N_{\alpha 1}$ and $N_{\alpha 2}$, on sublattices

$\alpha_1$ and $\alpha_2$, respectively ($N_{\alpha 1}$ and $N_{\alpha 2}$ are in fact equal to one half of unlike $V_2$ bonds on the corresponding $\alpha$ sublattices), as well as the numbers of chains of the same length, $N_{\alpha 1}(l)$ and $N_{\alpha 2}(l)$, for lengths ranging from $l=1$ to $l=300$. The ratios $N_{\alpha 1}(l)/N_{\alpha 1}$ and $N_{\alpha 2}(l)/N_{\alpha 2}$ were then equilibrated through the MC process and the so obtained values were finally assigned to $f_{\alpha 1}(l)$ and $f_{\alpha 2}(l)$. The MC calculations were performed using single-spin-flip Glauber dynamics, where the oxygen concentration $x$ is a functions of temperature $T$ and chemical potential $\mu$. We have studied lattices with periodic boundary conditions that consisted of 400x400 oxygen chain sites (O(1) sites, that split into two nonequivalent sublattices $\alpha_1$ and $\alpha_2$, in OII phase), and as many sites on $\beta$ sublattice (O(5) sites). One MC step included flipping of all 2X(400X400) lattice spins and one MC run (at a particular point $(x, T)$) typically consisted of $3 \cdot 10^4$ to $5 \cdot 10^4$ MC steps, where only every tenth was used to calculate chain length distributions $f_{\alpha 1}(l)$ and $f_{\alpha 2}(l)$, $l=1,2,\ldots,300$, and other relevant quantities (oxygen sublattice occupancies $x_1$ and $x_2$, 3-fold Cu fractions $n_1$ and $n_2$, etc.). At a certain number of points we have even used a really large number of MC steps, ranging from $10^5$ to $3 \cdot 10^5$.

We used (2) to calculate doping for substitution levels $b_1=0.1$, and $b_2=0.2$ at constant reduced temperature $\tau=k_BT/V_1=0.35=const$ that we estimated here to be referring to room temperature (RT). The obtained $x$ dependences of $p$ (shown by open squares and triangles in Figure 1) were then inserted into (1) to obtain corresponding $T_c(x)$'s (shown by the same symbols in Figure 2). The parameters $b_{eff}$ and $\chi$ were varied and their optimal values were obtained so to achieve the best agreement between calculated and experimental $T_c(x)$ dependences (filled symbols [2,5,6]). The $l_{cr}$ parameter was taken to be equal to 4, not only because it had recently been successfully used to explain emergence of 60K and 90K plateaus of the parent $YBa_2Cu_3O_{6+x}$ system [20], but also for it is in very good correlation with theoretical estimations [14,15]. Calculations were made applying Monte Carlo (MC) method to the ASYNNNI model to obtain length distributions $f_{\alpha 1}(l)$ and $f_{\alpha 2}(l)$, and fractions $n_1$ and $n_2$. The distributions were determined in each MC step as $N_{\alpha 1}(l)/N_{\alpha 1}$ and $N_{\alpha 2}(l)/N_{\alpha 2}$, for chain lengths $l=1,\ldots,300$, and subsequently equilibrated through the MC process ($N_{\alpha 1}(l)$ and $N_{\alpha 2}(l)$ denote numbers of chains of the same length $l$, on oxygen site sublattices $\alpha_1$ and $\alpha_2$, respectively, while $N_{\alpha 1}$ and $N_{\alpha 2}$ correspond to total numbers of chains). At all calculated points of $(x,\tau)$ space it was found that obtained values obey fairly well the expected behavior $f_i(l)=\omega_i(1-\omega_i)^{l-1}$, $i=\alpha_1, \alpha_2$, where $\omega_i$ is the inverse of the corresponding average chain length [21]. Such a behavior of distributions $f_i(l)$ ensures a relatively quick convergence of the sums in (2). The values of interaction constants $V_1>0$, $V_2<0$, and $V_3>0$, that define the ASYNNNI model, were taken to be those that were obtained from linear muffin-tin orbital (LMTO) method by Sterne and Wille [22] for the case of $YBaCu_3O_{6+x}$ system. As it can be seen from Figures 1 and 2 the calculated $p(x)$ and $T_c(x)$ dependences occur to be in a remarkable correlation with the experimentally obtained ones (shown by filled symbols) and obtained values of $\chi$ agree with expected values extracted from experimental data at $x\approx 1$ [2,6]. We also calculated chain contribution $p_{chain}(x)$ alone, at $\tau=0.45=const$, which reveals a clear flat section over the regime of OII phase (starting from $x\approx 0.5$) and slightly penetrates into the region of OI phase (Figure 1 - open circles) and the corresponding $T_c(x)$ is shown in Figure 2. As it can be seen, these results correlate very well with experimental $T_c(x)$ of Jorgensen *et al.* [4] (solid line in Figure 2) for the two-plateaus phenomenon in $YBa_2Cu_3O_{6+x}$ ($b=b_{eff}=0$).

We therefore believe that it is the same model of charge transfer mechanism given by (2) that accounts for the two-plateaus $T_c(x)$ in parent system $YBa_2Cu_3O_{6+x}$ as well as in the Ca doped $Y_{1-b}(Ca)_bBa_2Cu_3O_{6+x}$, with the RT attributed to $\tau=0.45$ in the former case versus $\tau\approx0.35$ in the later being the only distinction due to altered interactions caused by introduction of Ca. In what follows we are going to elucidate a bit more what might be a possible physical reasoning that is lying in the background of relating the RT, at least formally, to different $\tau$ values in two aforementioned cases ($b=0$ and $b\neq0$).

Although, for the $YBa_2Cu_3O_{6+x}$ material, the LMTO interactions have often been used as the input for calculations based on the ASYNNNI model, a consensus on the real magnitudes of these interactions is not yet achieved, but it seems that at least for the nearest neighbor O-O interaction $V_1$ there is some consensus that $V_1$ should be ranking around $\approx 6.9 mRy$ [18,22]. This fixes scaling between $T$ and $\tau$ at a rate $\Delta\tau\approx0.1\leftrightarrow\Delta T\approx100K$, so that RT would then correspond to $\tau\approx0.30$. Taking into account that magnitudes of the next-to-nearest neighbor $V_2$ and $V_3$ may well influence the location of the top of OII phase at $x\approx0.5$, the OI-to-OII phase transition at this composition would have been detected around $T\approx600K$, had the LMTO values been fully trustworthy, since theoretical calculations show that the OII top occurs at $\tau\approx0.58(9)$ [21]. However, according to experiments, OII top lies at $\approx125°C$ [11, 17] which implicates that the LMTO $V_2$ and $V_2$ are probably not correct (it should also be noted that even the authors of Reference [22] have subsequently suggested a modified values of $V_2$ and $V_3$ [18]). On the other hand, if the LMTO $V_1$ value is taken as reliable, it opens a way for a new strategy to estimate the $\tau$ value that the RT should be assigned to, by comparing the distances between the RT and the top of OII phase along two axes: $\tau$ and $T$. Thus, in case of $YBa_2Cu_3O_{6+x}$, $\Delta T\approx125°C$ would place RT at some point around $\tau\approx0.45$.

For the $Y_{1-b}(Ca)_bBa_2Cu_3O_{6+x}$ system it is reasonable to assume that the introduction of Ca would alter the effective pair wise interactions making it obvious that $\tau\approx0.45$ would not correspond to the RT any more. Before addressing this issue further, it is worthwhile to recapitalize some relevant topics that are tightly connected to the physics of the ASYNNNI model. Firstly, as our extended analysis of the model statistics shows [21], for given interactions $V_1$, $V_2$, and $V_3$ and at any $\tau=const$ below the top of OII phase, there always exists a well defined value of a cutoff parameter $l_{cr}$ (we call it *the optimal value*, $l_{opt}(\tau)$), so that for $l_{cr}=l_{opt}(\tau)$ the $p_{chain}(x)$ remains constant over the regime of OII phase (at $x>0.5$), with this constant portion slightly penetrating into the OI phase beyond the OI/OII second order phase transition point. Generally, the $l_{opt}(\tau)$ increases with temperature decrease so that for the LMTO interactions [22] we have found $l_{opt}(\tau=0.30)=12$, $l_{opt}(\tau=0.35)=8$, $l_{opt}(\tau=0.38)=7$, $l_{opt}(\tau=0.40)=6$, and $l_{opt}(\tau=0.45)=4$. Thus, at $\tau=const=0.45$ when $l_{cr}=l_{opt}(\tau)=4$, the calculated $p_{chain}(x)$ reveals a clearly visible plateau section (open triangles in Figure 1) and corresponding $T_c(x)$ displays prominent plateau at 60K accompanied with somewhat less distinctive plateau at 90K that is associated with underdoped-to-overdoped transition at $x\approx0.91$ as shown in Figure 2 (the $YBa_2Cu_3O_{6+x}$ case [20]). On the other hand, if $l_{cr}$ were chosen, at a given $\tau=const$, to be less than the corresponding $l_{opt}(\tau)$, the plateau would vanish to promote monotonously increasing $p_{chain}(x)$ with a characteristic change of slope at OII stoichiometry $x\approx0.5$. The same effect can be obtained by maintaining the same $l_{cr}$ but lowering the temperature, so that $l_{cr}$ becomes less than $l_{opt}(\tau)$ (since the later increases with reducing $\tau$). Thus, although the two calculated $p(x)$ in Figure 1, at $\tau=0.35$ ($4=l_{cr}<l_{opt}(\tau)=8$), increase in a monotonous fashion, they nevertheless reveal a characteristic kink feature at $x \approx 0.5$ (the OII stoichiometry) that is a remnant of the plateau that existed at $\tau=0.45$. Though not that prominent as those of calculated $p(x)$, the change of slope around $x \approx 0.5$ is also clearly visible in experimental $p(x)$ dependences (filled symbols). Secondly, not far below the top of OII phase almost all relevant thermodynamic quantities of the ASYNNNI model practically cease to be influenced by the magnitude of $V_3$, but they depend only on $V_2$ that emerges coupled to $T$ through the ratio $V_2/k_BT$. Some of the quantities scale with $\exp(-2|V_2|/k_BT)$ (fractions $n_1$ and $n_2$ of the 3-fold Cu, energy deviation $\Delta E$ from the ground state energy $E_0(x)$), the other scale with $\exp(2|V_2|/k_BT)$ (the average chain length and $l_{opt}(\tau)$), while some are $T$ and $V_i$-independent ($i=1,2,3$) but depend only on $x$ (oxygen sublattice occupations $x_1$ and $x_2$). The magnitude of $V_3$ can exert influence only on location of the top of OII phase along $\tau$ axes and on the upper limit $T_l$ of low temperature interval $0<T<T_l$ within which the ASYNNNI model turns to its $V_3$-independence.

The above facts, when put together with the lack of available experimental data on various aspects of phase dislocation in $(x,T)$ space in case of $Y_{1-b}(Ca)_bBa_2Cu_3O_{6+x}$ system (at different levels $b$ of Ca substitution) in particular, the position of the top of OII phase, make it obvious that at present stage there is a plenty of space for estimation of what the reduced temperature $\tau$ should the RT be assigned to. The effect of the altered $V_i$'s can be included by maintaining in fact the same LMTO values but shifting only $\tau$, because it is the $V_2$ that is really relevant and it enters into calculations coupled to $T$ through $V_2/k_BT$. It is, first of all, clear that $\tau$ referred to RT should be less than 0.45 since $l_{cr}=4<l_{opt}(\tau)$ ensures a monotonous increase of $p(x)$, for otherwise, if $l_{cr}$ were greater than $l_{opt}(\tau)$, the 60K plateau section of $p(x)$ would be transformed into a peak around $x \approx 0.5$ followed by a dip at $x>0.5$, before rising up again to its $\chi/2$ value at $x \approx 1$ [21]. By varying parameters $\chi$ and $b_{eff}$ to obtain a better fitting to the experimental $T_c(x)$'s we so arrived at a conclusion that the best correlation is achieved for $\tau=0.35$, and it is therefore our best estimate for RT for the case of $Y_{1-b}(Ca)_bBa_2Cu_3O_{6+x}$ ($b=0.1, 0.2$).

It is interesting to note that the so obtained values of $\chi$ (and $b_{eff}$) do not in fact appear to be strongly affected by $\tau$, at least, as long as $l_{cr}$ is sustained well below $l_{opt}(\tau)$. For example, at the substitution level $b_2=0.20$ and at $\tau=0.30$ ($l_{opt}(\tau)=12$), it was found that $\chi/2=0.165$, while at $\tau=0.38$ ($l_{opt}(\tau)=7$) it is $\chi/2=0.169$. Similarly, for the $b_1=0.1$, $\chi/2$ varies between 0.182 ($\tau=0.30$) and 0.189 ($\tau=0.38$). Therefore, the parameter $\chi$ is practically uniquely defined for it changes only slightly in a wide temperature interval around $\tau \approx 0.35$.

In summary, we have shown that there is a common charge transfer model, expressed by Equation (2), that lies behind doping mechanism and $x$ dependence of $T_c$ of both $Y_{1-b}(Ca)_bBa_2Cu_3O_{6+x}$ and $YBa_2Cu_3O_{6+x}$ compounds. A direct evidence that the chain contribution to doping is proportional to active chain hole concentration $h$ has been provided, $p_{chain}=(\chi/2)h$, while from $b$ dependence of averaged hole ability $\chi$ to attract electrons from $CuO_2$ planes, it has been indirectly shown that the $p_{chain}$ is clearly influenced by concentration $\rho_e$ of the available electrons in the planes.

In Figure 1, $x$ coordinates of the intersection points of the horizontal line, that corresponds to optimal doping $p=0.16$, with the calculated $p(x)$ dependences, explains why the maximum of $T_c$ shifts to lower oxygen concentrations as Ca content $b$ increases. Similarly, the way calculated $p(x)$ intersect horizontal line at critical doping $p_{crit}=0.19$, that is associated with the quantum critical point at which pseudogap phase vanishes, gives a clear indication of why the highly overdoped regime ($p>0.19$) is rather easily achievable in $Y_{1-b}(Ca)_bBa_2Cu_3O_{6+x}$, but so difficult to obtain in $YBa_2Cu_3O_{6+x}$.

## Acknowledgements

This work has been funded by Serbian Ministry of Science and Technology through the Project 141014. V. M. Matic is indebted to Dr Danka Savic for having thoroughly read the article and for the valuable comments made.

## Figure Captions

**Figure 1.** $x$ dependence of doping for $b=0$, $b=0.1$, $b=0.2$ shown by circles, triangles and squares, respectively. Open symbols show values calculated by (2) for $l_{cr}=4$, at $\tau=0.45=const$ for $b=0$ ($\chi=0.394$) and at $\tau=0.35=const$ for $b=0.1$ and $0.2$ ($\chi=0.370$, $b_{eff}=0.078$ and $\chi=0.334$, $b_{eff}=0.155$ respectively), while black symbols are experimental values scanned from References [2,5,6]. For $b=0.2$ case, the values of $p(x)$ (black squares) were obtained from experimental $T_c(x)$ [5,6] by using relation (1).

**Figure 2.** $T_c(x)$ for $b=0$, $b=0.1$, $b=0.2$ shown by circles triangles and squares, respectively. Open symbols show values obtained from the calculated $p(x)$ dependences (Figure 1) applying universal relation (1). Black squares denote experimental $T_c(x)$ values scanned from the Reference [6] ($b=0.2$). Black triangles stand for $T_c(x)$ values obtained from experimental $p(x)$ of Reference [2] ($b=0.1$) and solid line is scanned from the Reference [4] which stood there as a guide-to-eye line of their experimental $T_c(x)$ ($b=0$ case of the bare $YBa_2Cu_3O_{6+x}$ compound).

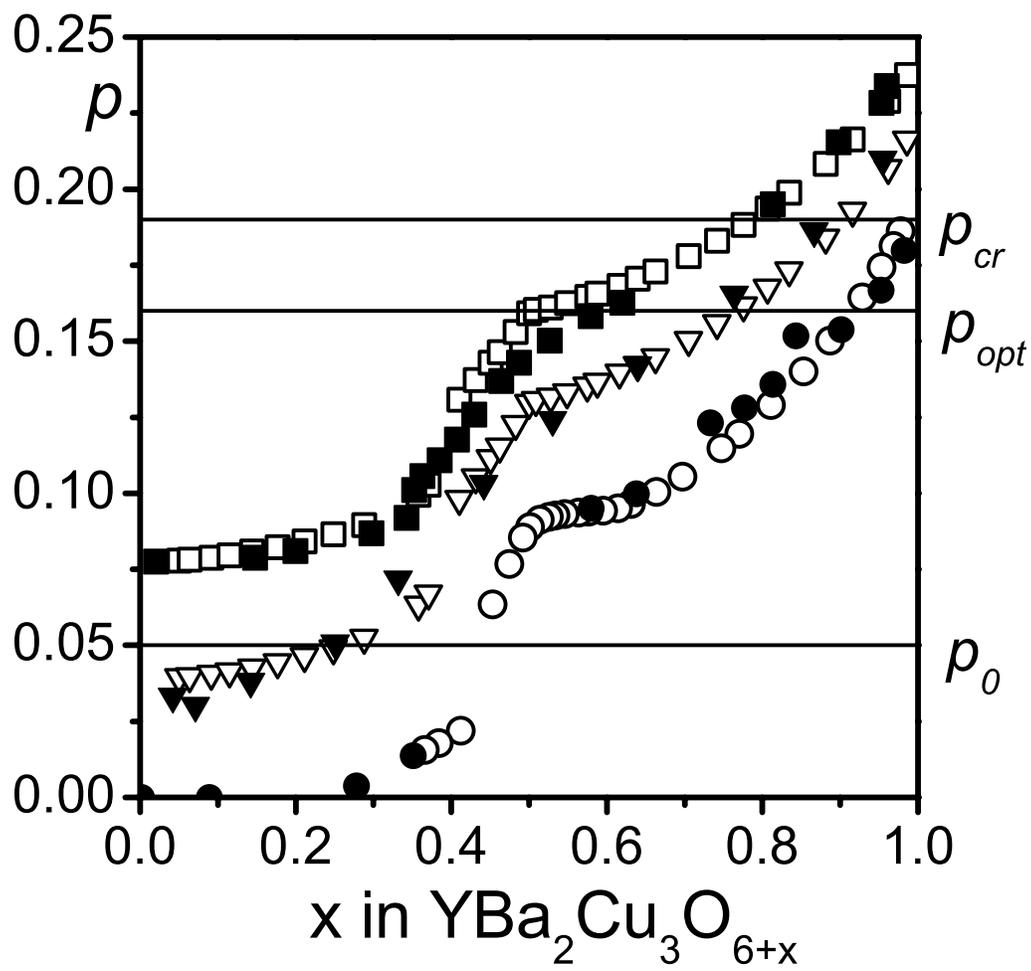

Figure 1

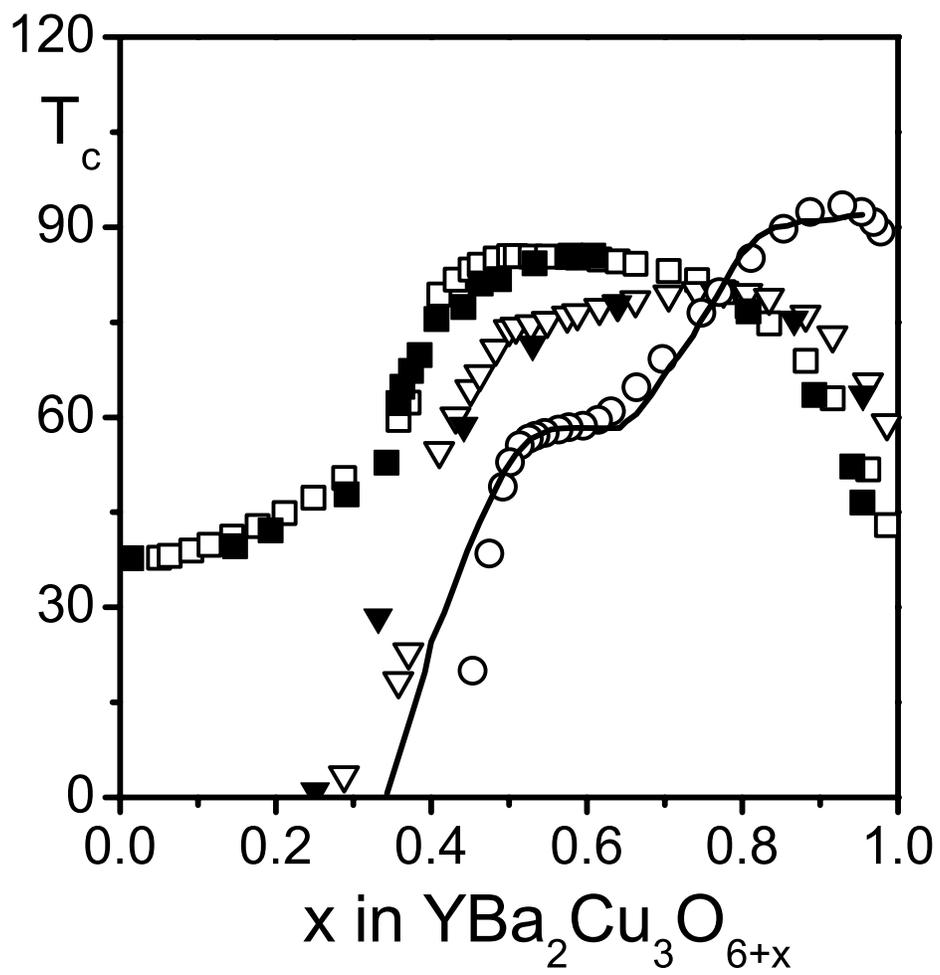

Figure 2